\documentclass[12pt,preprint]{aastex}
\usepackage{emulateapj5,epsf}

\shorttitle{Type I ULIRGs}
\shortauthors{Kawakatu et al.}

\begin{document}

\title{Type I ULIRGs: Transition Stage from ULIRGs to QSOs}


\author{Nozomu Kawakatu\altaffilmark{1}}
\affil{International School for Advanced Studies, Via Beirut 2-4, 34014 Trieste, Italy}

\author{Naohisa Anabuki\altaffilmark{2}}
\affil{Department of Earth and Space Science, Graduate School of Science, 
Osaka University, 1-1 Machikaneyama, Toyonaka, 560-0043 Osaka, Japan}

\author{Tohru Nagao\altaffilmark{3}}
\affil{Osservatorio Astrofisico di Arcetri, Largo Enrico Fermi, 5, 50125 Firenze, Italy \\ National Astronomical Observatory of Japan, 2-21-1 Osawa, 
Mitaka, Tokyo 151-8588, Japan}

\author{Masayuki Umemura\altaffilmark{4}}
\affil{Center for Computational Sciences, University of
  Tsukuba, Ten-nodai, 1-1-1 Tsukuba, Ibaraki, 305-8577, Japan}

\and

\author{Takao Nakagawa\altaffilmark{5}}
\affil{Institute of Space and Astronautical Science, Japan Aerospace Exploration Agency, JAXA, 3-1-1 Yoshinodai, Sagamihara, Kanagawa 229-8510, Japan}


\altaffiltext{1}{kawakatu@sissa.it}
\altaffiltext{2}{anabuki@ess.sci.osaka-u.ac.jp}
\altaffiltext{3}{tohru@arcetri.astro.it}
\altaffiltext{4}{umemura@ccs.tsukuba.ac.jp}
\altaffiltext{5}{nakagawa@ir.isas.jaxa.jp}


\begin{abstract}

We examine whether the ultraluminous infrared galaxies that contain a type I Seyfert nucleus (a type I ULIRG) are in the transition stage from ULIRGs to quasi-stellar objects (QSOs). To inspect this issue, we compare the black hole (BH) mass, the bulge luminosity and the far infrared luminosity among type I ULIRGs, QSOs and elliptical galaxies. As a result, we find the following results; (1) The type I ULIRGs have systematically smaller BH masses in spite of the comparable bulge luminosity relative to QSOs and elliptical galaxies. (2) The far-infrared luminosity of most type I ULIRGs is larger than the Eddington luminosity. We show that above results do not change significantly for 3 type I ULIRGs that we can estimate the visual extinction from the column density.
Also, for all 8 type I ULIRGs, we investigate the effect of uncertainties of BH mass measurments and our sample bias, so that it turns out that our results do not alter even if we consider above two effects.
In addition, Anabuki (2004) revealed that their X-ray properties are similar to those of the narrow line Seyfert 1 galaxies. These would indicate that active galactic nuclei (AGNs) with a high mass accretion rate exist in the type I ULIRGs.
Based on all of these findings, we conclude that it would be a natural interpretation that type I ULIRGs are the early phase of BH growth, namely the missing link between ULIRGs and QSOs. Moreover, by comparing our results with a theoretical model of a coevolution scenario of a QSO BH and a galactic bulge, we show clearly that this explanation would be valid.

\end{abstract}
\keywords{galaxies:active --- galaxies:bulges --- galaxies:formation --- galaxies:starburst --- quasars:general --- black hole}

\section{Introduction}
Up to now, the Infrared Astronomical Satellite ($IRAS$) made the remarkable discovery of a new class of galaxies, the ultraluminous infrared galaxies (ULIRGs, i.e., those having infrared luminosities greater than $L_{\rm IR}(8-1000\mu\,{\rm m})\geq 10^{12}L_{\odot}$), which emit the bulk of their energy at infrared wavelengths. 
A lot of studies (e.g., Sanders \& Mirabel 1996) have established that ULIRGs are gas-rich galaxies that in most cases have undergone a recent strong interaction with other galaxies eventually leading to a complete merger of the two. 
A commonly accepted explanation is that during such a galaxy collision 
the interstellar medium is transported toward the circumnuclear environment and is concentrated and compressed there, thus resulting in starburst on scales of less than a few kiloparsecs. 
In the next phase, the low angular momentum gas in the starburst region 
may fall into and accrete onto the central massive BH. 
In addition, the luminosity and space density of ULIRGs are similar to those of QSOs. Moreover, the luminosity function of IRAS galaxies is of a double power-low type, and thus differs from that of the normal galaxies that show exponential rollover and is rather similar to that of QSOs or starbursts (Scoville 1992). Thus, it has been suggested that ULIRGs are powered by the heavily obscured QSOs (e.g., Sanders et al. 1988) or the starbursts (Joseph \& Wright 1985). However, the physical relation between ULIRGs and QSOs has been an issue of long standing.

For the early type galaxies and QSOs, recent high spatial-resolution observations have suggested that the mass of a supermassive black hole (SMBH) tightly correlates with the mass, the velocity dispersion, and the luminosity of a galactic bulge (e.g., Kormendy \& Richstone 1995; Richstone et al. 1998; Laor 1998; Tremaine et al. 2001; McLure \& Dunlop 2001, 2002; Marconi \& Hunt 2003; Kawakatu \& Umemura 2004). It has been found that the relatively low-$z$ QSO hosts are mostly luminous and well-evolved early-type galaxies (e.g., McLeod \& Rieke 1995; Bahcall et al. 1997; Hooper, Impey, \& Foltz 1997; McLoed, Rieke \& Storrie-Lombardi 1999; Brotherton et al. 1999; Kirhaokos et al. 1999; McLure et al. 1999; McLure, Dunlope, \& Kukula 2000; Falomo et al. 2003; Dunlop et al. 2003). Recently, Veilleux, Sanders \& Kim (1999) have shown that the percentage of AGNs is $30-50\%$ for $L_{\rm IR} > 10^{12}L_{\odot}$. These findings suggest that the formation of a ULIRG, a SMBH, a galactic bulge and a QSO would be related to each other. However, this physical link is an open question.

From the theoretical points of view, Kawakatu, Umemura \& Mori (2003; hearafter KUM03) suggested a potential mechanism to build up a SMBH.
They consider the effect of the radiation drag \footnote{The radiation drag in the solar system is known as the Poynting-Robertson effect. Note that, in the early universe, Compton drag force has a similar effect on the formation of massive BHs (Umemura, Loeb, \& Turner 1993).}, which extracts angular momentum from interstellar medium in starburst galaxies and thereby drives the mass accretion onto a galactic center (Umemura 2001; Kawakatu \& Umemura 2002; Sato et al. 2004). On the basis of the radiation drag model, they proposed a new picture for a QSO formation. First, we regarded classical ULIRGs as the starburst galaxies, in which there is little AGN activity (if any). Next, we predicted the possibility of the ``proto-QSO phase", which is the optically thin and the total luminosity is dominated still by the bulge stars, although there is significant AGN activity in them. In this phase, a BH is still growing through the mass accretion and the BH-to-bulge mass ratio is smaller than that of QSOs and elliptical galaxies. And then, once a BH has grown fully and the central AGN dominates the total luminosity, the galaxy is regarded as a QSO (see also figures in KUM03). However, a proto-QSO has not been identified observationally yet although it is essential to clarify what objects correspond to proto-QSOs.

Recently, ultra deep X-ray observations suggested that the submillimeter galaxies (SMGs) at $z\simeq 1-3$ have the mass accretion rates approximately an order of magnitude lower than those of the coeval QSOs, assuming the Eddington luminosity (Alexander et al. 2005). Moreover, Borys et al. (2005) found that SMGs have smaller BH masses than QSOs with respect to the same mass range of bulges, and thus the SMGs may correspond to the ``proto-QSO" predicted in KUM03. In the nearby galaxies, Canalizo \& Stockton (2001) proposed that the infrared-selected type I AGNs (hereafter we refer them the ``type I ULIRGs'') are a transitional stage between ULIRGs and QSOs as their host galaxies are undergoing the tidal interactions or the mergers accompanied by the massive starbursts (see also Zheng et al. 2002; L\'{i}pari et al. 2005). 
In addition, most of them indicate the full width with half-maximum (FWHM) of the broad ${\rm H}{\rm \beta}$ line less than 2000 km/s (Zheng et al. 2002), thus AGNs would be actually narrow line Seyfert 1 galaxies (NLS1s). 
Recently, Anabuki (2004) studied 27 ULIRGs using the X-ray imaging and spectroscopic observations with {\it ASCA}, {\it Chandra}, and {\it XMM-Newton}. Among their sample, 10 ULIRGs were identified as the type I ULIRGs. 
After correcting the absorption effects, he found that seven luminous type I ULIRGs show the soft X-ray excess, the SED with a steep photon index ($\Gamma_{2-10{\rm kev}} > 2$), and also the violent flux change (excess variance $\sim 0.01-0.1$), which are characteristic properties of NLS1s. These X-ray properties would imply that AGNs with smaller BHs and high mass accretion rates exist in type I ULIRGs (e.g., Pound et al. 1995; Boller et al. 1996; Mineshige et al. 2000). Moreover, Mathur et al. (2000) suggested that NLS1s may be Sy1s in the early stage of their evolution if the BHs in NLS1s are under massive with respect to their host bulges. Hence, it is likely that all of above characteristics support the hypothesis that the type I ULIRGs are the early phase of BH growth. 
However, the previous works have never examined that the BH-to-bulge relation among type I ULIRGs, QSOs and elliptical galaxies. By investigating this issue, we can reveal whether type I ULIRGs have systematically smaller BH than QSOs and elliptical galaxies. By combining the BH-bulge relation for type I ULIRGs with the previous works for type I ULIRGs, we will test if they are really ``proto-QSOs", which are the transition phase from ULIRGs into QSOs. To this end, we demonstrate the relation among type I ULIRGs, QSOs, elliptical galaxies on a BH mass ($M_{\rm BH}$) versus an absolute {\it R}-band magnitude of a galactic bulge ($M_{\rm R}(\rm bulge)$). Also, we examine a BH mass ($M_{\rm BH}$) and a far-infrared (FIR) luminosity ($L_{\rm FIR}$) for type I ULIRGs and QSOs, in order to constrain the origin of type I ULIRGs.

The paper is organized as follows. 
In $\S2$, we describe how type I ULIRGs, QSOs and elliptical galaxies are selected.
In $\S3$, we briefly review the technique of estimating BH 
masses from the broad emission-line widths for type I ULIRGs.  
In $\S4$, we plot the data of type I ULIRGs in the $M_{\rm BH}-M_{\rm R}(\rm bulge)$ diagram and also compare them with that of QSOs and elliptical galaxies. Next, we show the $M_{\rm BH}-L_{\rm FIR}$ elation for type I ULIRGs and QSOs, and then compare with each others. 
Finally, we constrain the optical extinction of the central regions for type I ULIRGs by using the results of hard X-ray observations.
In $\S5$, we summarize our observational results for the type I ULIRGs, and then we compare them with one of the theoretical models (KUM03 model). 
Section 6 is devoted to the conclusions. 
 Through this paper, we adopt the Hubble parameter $H_{0}$=75 km ${\rm s}^{-1}$ ${\rm Mpc}^{-1}$ and the deceleration parameter $q_{0}$=0.5, and have converted the results from published papers to this cosmology to facilitate with comparisons.

\section{Sample Selection}
The aim of our study is to clear if type I ULIRGs are really the transition phase from ULIRGs into QSOs. To accomplish this, we need to use a type I ULIRG sample as the data of FWHM (${\rm H}\beta$), the optical luminosity at $5100{\rm \AA}$ in the rest frame and the {\it R}-band absolute magnitude of host bulges are coeval in these objects. For comparison, we also compile an optically-selected QSO sample and an elliptical galaxy sample, for which the BH mass and the {\it R}-band absolute magnitude of a bulge are available. The details of these samples are given as follows. 

(1) The type I ULIRG sample is from Zheng et al. (2002). This sample was compiled from ULIRGs in the QDOT redshift survey (Lawrence et al. 1999), the 1Jy ULIRG survey (Kim \& Sanders 1988), and an IR QSO sample selected from the cross-correlation of IRAS Point-Source Catalogue with the ROSAT All-Sky Survey catalog (Boller et al. 1992). All the type I ULIRGs selected by Zheng et al. (2002) are ULIRGs with mid-infrared to far-infrared properties from IRAS observations. From this sample, we choose all 8 type I ULIRGs (IRAS F07599+6508, IRAS F11119+3257, IRAS Z11598-0112, IRAS F13342+3932, IRAS F15462-0450, IRAS F21219-1757, Mrk 231, and Mrk 1014), for which have both data of the width of broad ${\rm H}\beta$ line, the luminosity at $5100{\rm \AA}$ and the $R$-band absolute magnitude of host galaxies, $M_{\rm R}(\rm bulge)$. Therefore, these objects are the best and maximal sample to achieve our aim at this time. As for 8 type I ULIRGs, we obtain the FWHM (${\rm H}\beta$) and the optical luminosity at $5100{\rm \AA}$ in the rest frame by Zheng et al. (2002) and $M_{\rm R}(\rm bulge)$ by Veilleux et al. (2002). In their paper, the contribution from the $R$-band absolute magnitude of AGNs were removed for $M_{\rm R}(\rm bulge)$ (see Veilleux et al. 2002 for this procedure). In addition, according to Veilleux et al. 2002, all 8 type I ULIRGs are the single nucleus onjects. In Zheng sample, 23/25 type I ULIRGs have the optical luminosity at $5100{\rm \AA}$ and the width of broad ${\rm H}{\beta}$ line. Then, we have compared the 8 selected objects with the rest 15 objects for these two properties. Figure 1 shows the optical luminosity at $5100{\rm \AA}$ against the width of broad ${\rm H}{\beta}$ line. The filled red circles denote the 8 selected type I ULIRGs, while the open red circles represent the rest 15 type I ULIRGs whose bulge luminosities were not available. As seen in this figure, the 8 selected type I ULIRGs would be representative of large population  with respect to the optical luminosity at $5100{\rm \AA}$ and the width of broad ${\rm H}{\beta}$ line. In addition, the significant differences between our sample and the rest does not appear for the rages of the redshift ($0.1 < z < 0.4$) and infrared luminosity ($L_{\rm IR} > 10^{12}L_{\odot}$).

(2) The optically-selected QSO sample comprises 29 Palomar Green quasars (PG QSOs) from 30 luminous quasars ($M_{\rm V} < -23$) published by McLure \& Dunlop (2001). This QSO sample consists of two optically matched subsamples of 17 radio-quiet QSOs and 13 radio-loud QSOs. The advantage of this sample is that all members have accurate bulge luminosities available from two-dimensional modeling of {\it HST} images. The average redshift of the QSO sample is around 0.2. In this paper, we excluded PG 0157+001 (Mrk 1014) from the optically-selected QSO sample since it is categorized by type I ULIRGs. All 29 QSOs have the data of a BH mass and {\it R}-band magnitude of a bulge compiled by McLure \& Dunlop (2001) and Dunlop et al. (2003). In their paper, the contribution from the $R$-band absolute magnitude of AGNs were also removed for $M_{\rm R}(\rm bulge)$. For 13 PG QSOs in our sample, their infrared flux was taken from Sanders et al. (1989) and Haas et al. (2000, 2003).  

(3) The elliptical galaxy sample consists of 20 objects drawn from the list of 37 nearby inactive galaxies with dynamical BH measurements published by Kormendy \& Gebhardt (2002). In this paper, our main purpose is to investigate the physical link between type I ULIRGs, QSOs and elliptical galaxies. Thus, we excluded those galaxies in the Kormendy \& Dunlop list that were not E-type morphology (including lenticulars). The Kormendy \& Gebhardt list is made up of 20 E-type galaxies. All 20 elliptical galaxies have the data of the BH mass and the {\it B}-band absolute magnitude of the bulge (Kormendy \& Gebhardt 2002; Gebhardt 2003). To convert the {\it B}-band magnitude to the {\it R}-band, standard bulge colors of {\it B}-{\it R}=1.57 were assumed (Fukugita et al. 1995).

We summarized the various physical parameters of type I ULIRGs, QSOs and 
elliptical galaxies in Table 1.

\vspace{5mm}
\epsfxsize=8cm 
\epsfbox{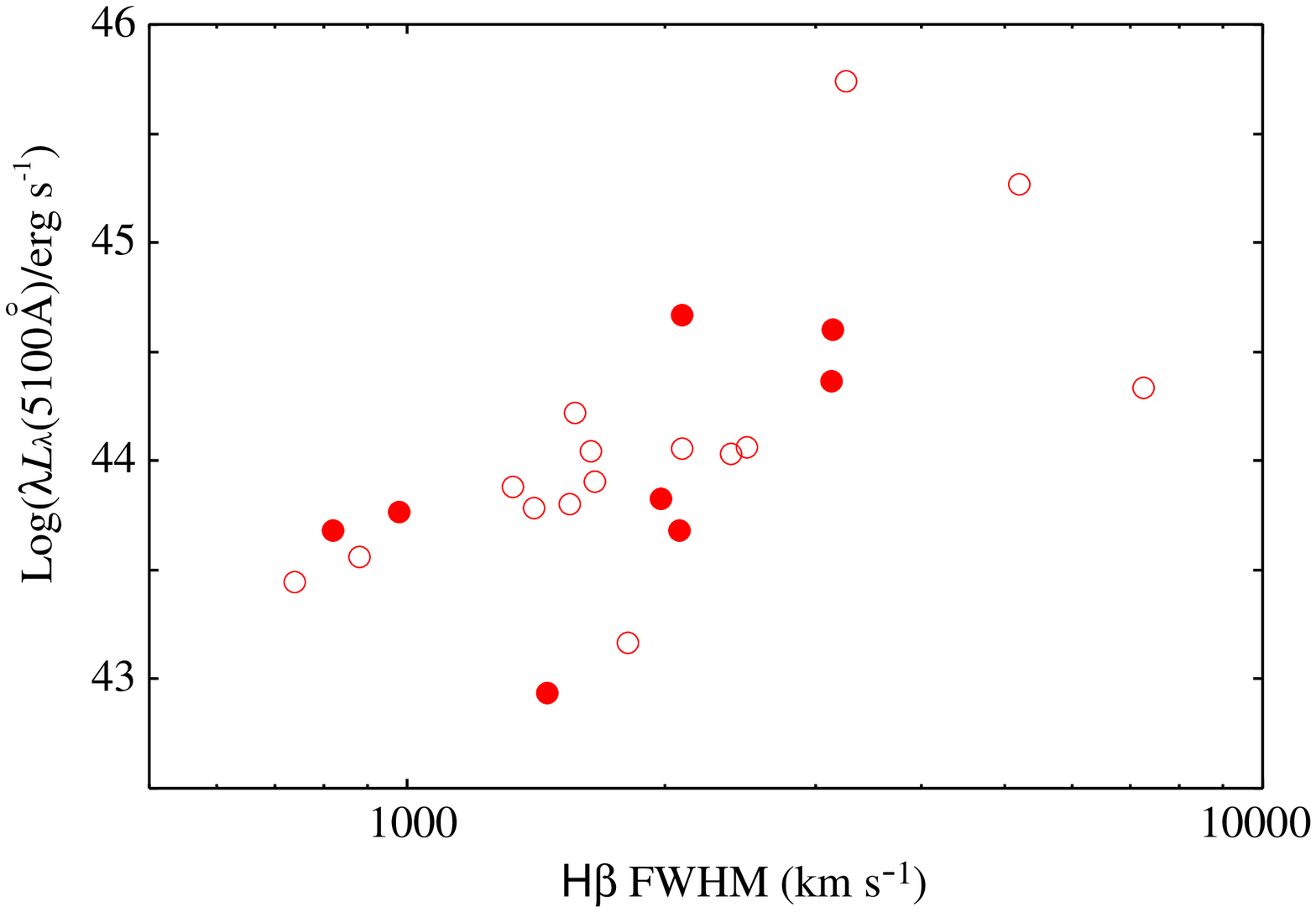}
\figcaption
{
Optical luminosity at $5100{\rm \AA}$ against the width of broad ${\rm H}{\beta}$ line. The filled red circles denote the 8 selected type I ULIRGs, while the open red circles represent the rest 15 type I ULIRGs. The distribution of the 8 selected type I ULIRGs are similar to that of the rest.
}

\section{Black Hole Estimate of Type I ULIRGs} 

As mentioned $\S 2$, the BH masses of QSOs and elliptical galaxies have been already estimated by previous works, but we do not have those of 8 type I ULIRGs. Thus, we need to estimate them within the present paper. 
The method to estimate a BH mass is based on the assumption that the motion of ionized gas clouds moving around the black hole is dominated by  the gravitational force and the clouds within the broad line region (BLR) is virialized (e.g., Peterson \& Wandel 1999, 2000). Thus, the central black hole mass can be expressed by $M_{\rm BH}\approx R_{\rm BLR}v^{2}/G$, where $v$ is the velocity dispersion of matter at the size of the broad line region $R_{\rm BLR}$, which is gravitationally bound to the BH. 
Then, the central mass can be estimated as 

\begin{equation}
M_{\rm BH}=1.5\times 10^{5}\left(\frac{R_{\rm BLR}}{\rm lt-days}\right)
\left(\frac{v_{\rm FWHM}}{10^{3}{\rm km}\,{\rm s}^{-1}}\right)^{2}M_{\odot}.
\end{equation}

The velocity dispersion $v$ can be estimated from the FWHM of ${\rm H}\beta$ broad line emission $v=fv_{\rm FWHM}$, by assuming the BLR gas is in isotropic motions ($f={\sqrt{3}}/2$).
Based on 17 Seyfert galaxies and 17 optically-selected PG QSOs, Kaspi et al. (2000) determine an empirical relationship between the size of the broad-line region, $R_{\rm BLR}$, and optical continuum luminosity, $\lambda L_{\lambda}(5100{\rm \AA})_{\rm rest}$, where $R_{\rm BLR}$ is the distance of the emission-line clouds responding to the central continuum variation as determined from reverberation mappings (see Kaspi et al. 2000): 

\begin{equation}
R_{\rm BLR}=(32.9^{+2.0}_{-1.9})\left[\frac{\lambda L_{\lambda}(5100{\rm \AA})_{\rm rest}}{10^{44}{\rm erg}\,{\rm s}^{-1}}\right]^{0.70\pm 0.033} {\rm light}-{\rm days}.
\end{equation}

By combining the equation (1) and (2), we can obtain the following formula:

\begin{eqnarray}
M_{\rm BH}=(4.9^{+0.4}_{-0.3})\times 10^{6}\left[\frac{\lambda L_{\lambda}(5100{\rm \AA})_{\rm rest}}{10^{44}{\rm erg}\,{\rm s}^{-1}}\right]^{0.70\pm 0.033}\nonumber \\
\left(\frac{v_{\rm FWHM}}{10^{3}{\rm km}\,{\rm s}^{-1}}\right)^{2}M_{\odot}.
\end{eqnarray}
As seen from the equation (3), the luminosity at $5100{\rm \AA}$ in the rest frame and the FWHM of ${\rm H}{\rm \beta}$ are needed to estimate a BH mass. We should also keep in mind that the equation (2) holds not only for broad-line type I AGNs, but also for NLS1s (Peterson et al. 2000). 
In order to evaluate the optical luminosity at $5100{\rm \AA}$ in the rest frame, $L_{\lambda}(5100{\rm \AA})_{\rm rest}$, we use the formula $L_{\lambda}(5100{\rm \AA})_{\rm rest}=4\pi d_{\rm L}^{2}(1+z)F_{\lambda}(5100(1+z){\rm \AA})_{\rm obs}$, where $d_{\rm L}$ is the luminosity distance which is given by $d_{\rm L}=(cz/H_{0})(1+z/4)$ for a small value of $z$. Here, the FWHM (${\rm H}{\rm \beta}$) and the observed flux at $5100{\rm \AA}$, $F_{\lambda}(5100(1+z){\rm \AA})_{\rm obs}$, are given by Zheng et al. (2002) and are measured directly from their spectra.

In this paper, we assume that the contribution from central AGNs dominate the optical emission at $5100{\rm \AA}$ in the rest frame and that the stellar continuum emission is negligible. To confirm if this assumption is reasonable, we check the flux ratio of the optical emission at $5100(1+z){\rm \AA}$ and the hard X-ray emission of 2-10(1+$z$)keV for 5 type I ULIRGs. We get the flux of hard X-ray from Anabuki (2004). As a consequence, 3 objects (IRAS F11119+3257, IRAS Z11598-0112, and Mrk 1014) have nearly same flux ratio as that of PG QSOs, $F_{2-10(1+z){\rm keV}}/{F_{5100(1+z){\rm \AA}}}$=$10^{3-5}$. As for PG QSOs, we obtain this flux ratio from the data of $F_{2-10(1+z){\rm keV}}$ (George et al. 2000) and $F_{{5100(1+z){\rm \AA}}}$ (Kaspi et al. 2000). 
Although other 2 objects (IRAS F07599+6508 and Mrk 231), which are the broad absorption line QSOs (BAL QSOs), have an extremely low luminosity at the X-ray band, it has been considered that it is not be due to the intrinsic effect but the absorption effect (e.g., Gallagher et al. 2002). Thus, it is expected that these 2 BAL QSOs have almost same flux ratio as that of PG QSOs intrinsically. Since the contributions from AGNs dominate the optical emission for PG QSOs, it is valid assumption that the central AGNs in type I ULIRGs dominantly power the optical emission at $5100{\rm \AA}$ in the rest frame. All the basic parameters of 8 type I ULIRGs are listed in Table 2.

\section{Results}
\subsection{$M_{\rm BH}-M_{\rm R}$ Relation}
Figure 2 plots the $R$-band absolute magnitude of bulge components (spheroidal components), $M_{\rm R}\,(\rm bulge)[\rm mag]$, versus the black hole mass, $M_{\rm BH}\,[M_{\odot}]$ for 8 type I ULIRGs, 29 QSOs and 20 elliptical galaxies. In this figure, the red circles represent type I ULIRGs, the blue squares show QSOs, and the green circles denote elliptical galaxies. The solid line is the best-fitting relation for QSOs. This relation is given by $\log{(M_{\rm BH}/M_{\odot})}=-0.61(\pm 0.08)M_{\rm R}(\rm mag)-5.47(\pm 1.82)$, which is comparable to that found by Laor (1998). The underlines below the name of objects denote that they have the similar properties at the soft and the hard X-ray band as those of NLS1s (Anabuki 2004). As for all QSOs and ellipticals, they corrected the effect of Galactic and internal extinction (McLure \& Dunlop 2001; Dunlop et al. 2003; Gebhardt et al. 2003; Kormendy \& Gebhardt 2003). In the case of type I ULIRGs, the effect of the visual extinction may not be negligible because of significantly larger FIR luminosity, which would imply the plenty of dusty gas. Thus, the optical extinction may affect on the estimation of the BH masses and the bulge luminosities. Then, we exhibit the extinction effect toward the BLRs of type I ULIRGs as the arrows in Figure 2. Hereafter, we use $A_{\rm V}$ as the total extinction toward the BLRs of type I ULIRGs. Also, the effect of the optical extinction for the hosts of type I ULIRGs may be also significant (e.g., Veilleux et al. 2002). If this is the case, then we would underestimate the bulge luminosity of type I ULIRGs. 
As for the morphology of host galaxies, a lot of works have showed that the surface brightness profiles of QSO hosts resemble the de Vaucouleurs profile (e.g., McLure et al. 1999). As for 8 type I ULIRGs, the surface brightness profiles of 6 host galaxies in type I ULIRGs can fit with the de Vaucouleurs profile (elliptical-like) for $R$ and $K'$-band, and the others (IRAS 11598 and IRAS F15462) can fit with both the de Vaucouleurs profile and exponential profile (disk-like) for their band (Veilleux, Kim \& Sanders 2002). Thus, we put on the label ``(E/D)"  for the latter 2 objects in Figure 2.  In this paper, we focus on their absolute magnitude of spheroidal components. If the magnitude of galaxies are dominated by disk components, our 
estimations of the $R$-band absolute magnitude in bulges would be overestimated. Note that we assume $f=\sqrt{3}/2$ to estimate the BH mass for QSOs, while Mclure \& Dunlop used $f=3/2$, and thus our estimation for QSO BHs is three times as small as their estimation. 

As seen in Figure 2, we have found that type I ULIRGs have systematically smaller BH mass than QSOs and elliptical galaxies in spite of the comparable bulge luminosity, if the visual extinction effect is small ($A_{\rm V} < 3$) for type I ULIRGs. Namely, the BH mass ranges are $M_{\rm BH}\approx 10^{6-8}M_{\odot}$ for type I ULIRGs and $M_{\rm BH}\approx 10^{8-9}M_{\odot}$ for QSOs and elliptical galaxies. However, if $A_{\rm V}$ of all type I ULIRGs is larger than $\sim 3$, the BH mass of type I ULIRGs would be similar to that of QSOs. 
Thus, in order to justify if a BH mass of type I ULIRGs is systematically small, we need to constrain the effect of the visual extinction, which  will be discussed in $\S4.3$. On the other hands, if the extinctions for their host galaxies are significant, then the data points just move on top in Figure 2. In short, this effect makes the difference between the BH-mass distributions of type I ULIRGs, QSOs and ellipticals large. In addition, we find that the elliptical galaxies are located at slightly lower parts than QSOs in Figure 2. According to KUM03, this result may indicate that the hosts of QSOs are slightly younger than that of elliptical galaxies (for details, see $\S5$).

\vspace{5mm}
\epsfxsize=8cm 
\epsfbox{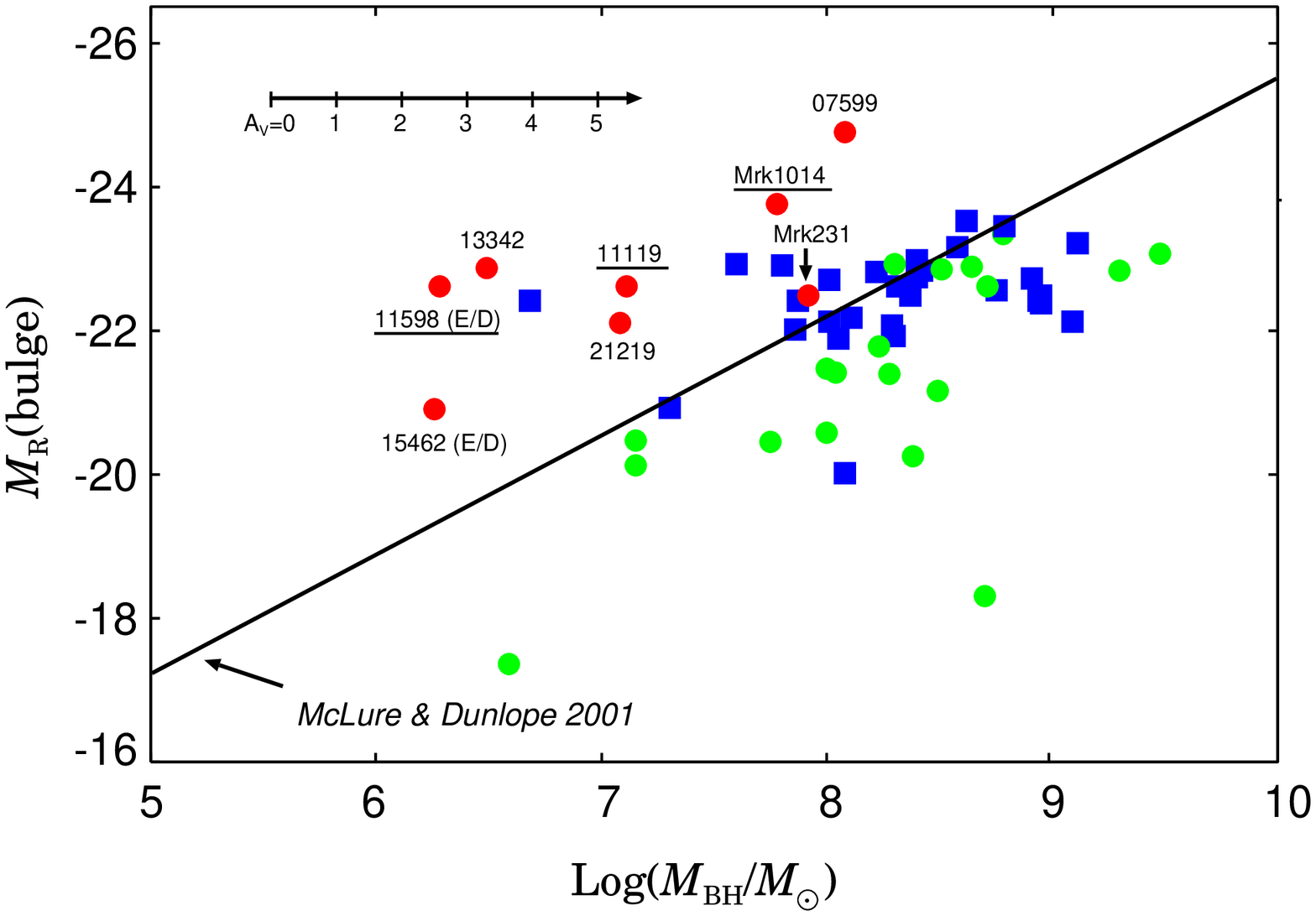}
\figcaption
{
Absolute $R$-band bulge magnitude versus black hole mass for 8 type I ULIRGs (red circles), 29 QSOs (blue squares) and 20 elliptical galaxies (green circles). The black hole masses for type I ULIRGs are derived from their broad H$\beta$ line widths by using equation (3). 
The black hole masses for QSOs are given by McLure \& Dunlop (2001) and Dunlop et al. (2003). 
The black hole masses for elliptical galaxies are the dynamical estimates as Kormendy \& Gebhardt (2003) and Gebhardt et al (2003). The black horizontal arrow shows the extinction effects for the BLR of type I ULIRGs. The under line below the name of objects denote that they have the same properties at hard X-ray band like NLS1s (Anabuki 2004).
The solid line is the best-fitting relation for the optical QSOs, which is $\log{(M_{\rm BH}/M_{\odot})}=-0.61(\pm 0.08)M_{\rm R}(\rm mag)-5.47(\pm 1.82)$.
The symbols ``(E/D)" denote the objects can fit with both de Vaucouleurs profile (elliptical-like) and exponential 
profile (disk-like) for $R$ and $K'$ band (Veilleux et al. 2002).
}
\vspace{5mm}

Finally, we check the effects of systematic errors in the methods used to evaluate BH masses and the effect of our sample bias. First, we discuss the systematic errors in methods. As for the type I ULIRGs, the uncertainties of the BH mass were estimated by error propagation using the optical luminosity at $5100 {\rm \AA}$ and FWHM of ${\rm H}{\beta}$ measurements given by Zheng et al. (2002). The mean error of the BH mass is a factor 1.3. Although McLure \& Dunlop (2001) did not show the uncertainties of BH masses clearly, in general the BH mass in the this way (see $\S 3$) is accurate within a factor 2-3 (e.g., Wang \& Lu 2001; Marziani et al. 2003; Shemmer et al. 2004). 
As seen in Figure 2, except for Mrk231, the BH mass of type I ULIRGs are ten times as small as that of QSOs and elliptical galaxies at the fixed {\it R}-band magnitude  of bulge. Furthermore, the mean error of BH mass for elliptical galaxies is a factor 3. Thus, it would be clear that the systematic difference of BH masses between type I ULIRGs and QSOs (or ellipticals) can not be explained by only systematic errors in methods of BH mass measurements. On the other hands, as shown in $\S 2$, the 8 selected type I ULIRGs would be representative of large population. Thus, the rage of BH masses in other 12 type I ULIRGs are similar to that of the 8 selected type I ULIRGs with $M_{\rm BH}=10^{6-8}M_{\odot}$. The rest three objects have massive BHs with $\approx 10^{9}M_{\odot}$. As a consequence, 12 type I ULIRGs would be located farthest from the location of QSOs if the range of {\it R}-band magnitude of bulge is from -22 to -24 that is typical range for the 8 selected type I ULIRGs. Thus, we found that the systematic difference we found in $M_{\rm R}(\rm bulge)-M_{\rm BH}$ diagram would not be the effect of our sample bias. We should keep in mind that othee results we will show later (Figure 2 and 3) also do not change significantly by the systematic errors in the BH measurements and the effect of our sample bias.

\subsection{$M_{\rm BH}$-$L_{\rm FIR}$ Relation}
Figure 3 shows that the BH mass-to-FIR luminosity relation,  $M_{\rm BH}-L_{\rm FIR}[40-500\mu{\rm m}]$ using the derived BH mass (see Table 1) and the FIR luminosity for 8 type I ULIRGs (Zheng et al. 2002).  In order to compare the result of type I ULIRGs with that of QSOs, we select 13 PG QSOs from the sample selected by McLure \& Dunlop (2001) and Dunlop et al. (2003). All 13 PG QSOs have the data of the IRAS flux densities at 60 and 100$\mu$m (Sanders et al. 1989; Haas et al. 2000, 2003). As for 13 PG QSOs, we calculated their far-infrared luminosities following formula (Sanders \& Mirabel 1996) ,based on the flux densities from the IRAS Faint Source Catalog: $L_{\rm FIR}[40-500\mu{\rm m}]=4\pi d_{\rm L}CF_{\rm FIR}$, where the scale factor $C(=1.4-1.8)$ is the correction factor required to account principally for extrapolated flux longward of the IRAS 100 $\mu$m filter, and $F_{\rm FIR}$ is defined as 
$1.26\times 10^{-14}\times (2.58f_{60}+f_{100})[{\rm W}\,{\rm m}^{-2}]$ with $f_{60}$ and $f_{100}$ being the IRAS flux densities at 60 and 100 $\mu$m in unit of Jy. Here, we employ $C=1.8$. The red circles show the type I ULIRGs and the blue squares and arrows represent QSOs.
The black horizontal arrow shows the optical extinction effect for the BLRs of type I ULIRGs. The thick solid line denotes the luminosity ratio of far infrared to Eddington luminosity ($L_{\rm FIR}/L_{\rm Edd}$) equals unity. The thin solid, dashed and dot-dashed line are $L_{\rm FIR}/L_{\rm Edd}=10$, $L_{\rm FIR}/L_{\rm Edd}=0.1$ and $L_{\rm FIR}/L_{\rm Edd}=0.01$, respectively.

\vspace{5mm}
\epsfxsize=8cm 
\epsfbox{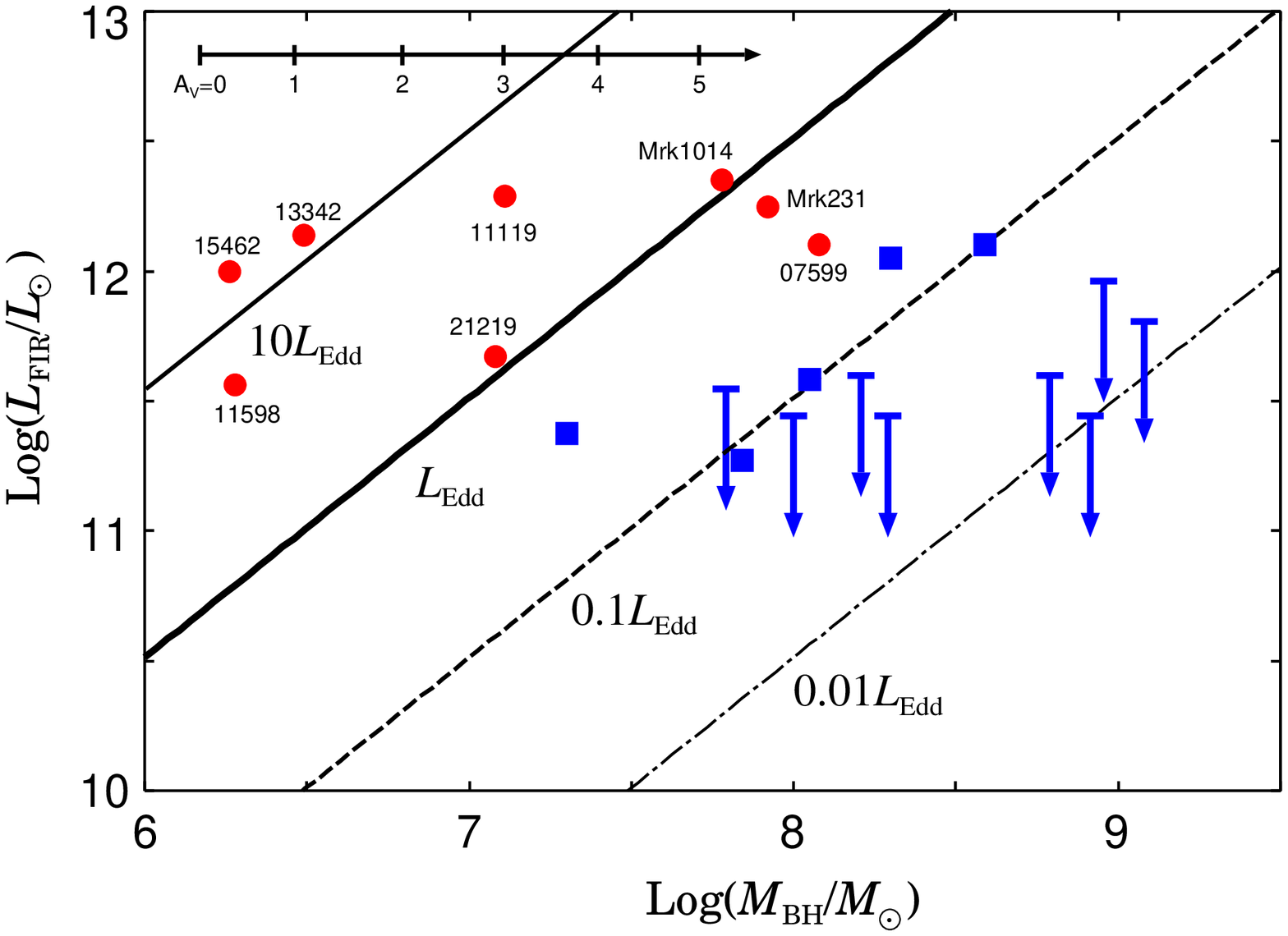}
\figcaption
{
Far-infrared luminosity versus the black hole mass for 8 type I ULIRGs (red circles) and 13 PGQSOs (blue squares and arrows). 
The black horizontal arrow shows the effect of visual extinction for the BLRs of type I ULIRGs. The thick solid line denotes the luminosity ratio of far-infrared to Eddington luminosity ($L_{\rm FIR}/L_{\rm Edd}$) equals unity. The thin solid, dashed and dot-dashed line are  $L_{\rm FIR}/L_{\rm Edd}=10$, $L_{\rm FIR}/L_{\rm Edd}=0.1$ and $L_{\rm FIR}/L_{\rm Edd}=0.01$, respectively. 
}
\vspace{5mm}

As seen in Figure 3, it turns out that the FIR luminosity is larger than the Eddington luminosity for most type I ULIRGs in Figure 3. By contrast, the FIR luminosity is more than one order of magnitude smaller than Eddington luminosity for QSOs, namely $L_{\rm FIR} < 0.1L_{\rm Edd}$. 
This may indicate AGNs in type I ULIRGs with a high mass accretion rate, or the existence of another power source, for which the promising candidate is a starburst. On this issue, we will discuss $\S5.1$. 
In addition, Figure 3 shows that type I ULIRGs have systematically a smaller BH mass than QSOs at fixed the FIR luminosity. On the other hands, the large far-infrared luminosity would imply the plenty of dusty gas (e.g., Haas et al. 2003). Hence, for type I ULIRGs it would show that the mass ratios of the total dusty gas to the BH are much larger than that of QSOs. It would indicate that BH masses of type I ULIRGs have growth potential. 

\subsection{Optical Extinction Effect}
As mentioned in $\S 4.1$ and $\S 4.2$, we found that type I ULIRGs have systematically smaller BHs than that of QSOs at given the absolute magnitude of bulge and given the FIR luminosity. However, the BH masses of type I ULIRGs would depend on the optical extinction effect, namely, the heavily extinction makes the derived BH mass smaller (see the horizontal arrow in Figure 2 and 3). On the other hands, it is well-known that Balmer decrement is a standard evaluation of optical extinction of the narrow line regions (e.g., Osterbrock 1989). However, this method is generally invalid to estimate the amount of the visual extinction toward BLRs, because the flux ratios of broad Balmer lines are sometimes seriously affected by collisional excitation effects. 
As a complementary approach, the hard X-ray observations enable us to make a measurement of the extinction toward the nucleus, and to constrain the BH mass more reliably. We should note that many observations suggested that the absorption column density ($N_{\rm H}$) derived from hard X-ray is systematically large relative to optical extinction of $A_{\rm V}$ under the assumption of Galactic gas/dust mass ratio (e.g., Maiolino et al. 2001a; Watanabe et al. 2004). 
Thus, we estimate the optical extinction of 3 type I ULIRGs (IRAS F11119+3257, IRAS Z11598-0112 and Mrk1014) with the relation $A_{\rm V}/N_{\rm H}=4.8^{+14.1}_{-3.6}\times 10^{-23} \,{\rm mag\, cm^{-2}}$ ($1\sigma$ dispersion), which is derived from the AGN observations (Maiolino et al. 2001b). 
As for two BAL QSOs (IRAS F07599+6508 and Mrk 231), the $A_{\rm V}/N_{\rm H}$ relation for normal QSOs would not hold on BAL QSOs. However, recent works  have found that the optical extinction of BAL QSOs is around 0.1-1 by comparing the composite non-BAL QSO spectra with the composite BAL QSO spectra (e.g., Brotherton et al. 2001; Richard et al. 2003). Therefore, the optical extinction of two BAL QSO in our sample may be less than $A_{\rm V}=1$. We summarize the visual extinctions for type I ULIRGs in Table 3. 
As for at least 3 type I ULIRGs to be able to estimate the visual extinction from the column density, we show that our results ($\S 4.1$ and $\S 4.2$) do not change drastically. 

\subsection{Summary of observational results}
Based on the above results, we summarize our findings on type I ULIRGs as follows. 
(1) The BH mass of type I ULIRGs is systematically smaller than QSOs and elliptical galaxies despite of the comparable bulge luminosity to them. 
(2) Most type I ULIRGs have particularly large FIR luminosity against the Eddington luminosity. We show that above results do not change significantly for 3 type I ULIRGs even if we consider the effects of the visual extinction. 
Also, for 8 type I ULIRGs, we investigate the effect of uncertainty of BH mass measurements and our sample bias, so that we found that our results do not alter. Additionally, their X-ray luminosity properties are similar to those of NLS1s, whose X-ray properties reflect a high mass accretion rate (Anabuki 2004 in details).
From all these findings, it would be a natural explanation that type I ULIRGs are the early phase of BH growth, namely the transition stage from ULIRGs to QSOs. In the next section, we will investigate whether this interpretation is reasonable by comparing our results with the theoretical prediction.

\section{Discussions}
Based on our observational results for 8 type I ULIRGs, we have suggested that type I ULIRGs would be the transition stage from ULIRGs to QSOs. However, it has not been cleared that the physical relationship between type I ULIRGs, QSOs and elliptical galaxies with the objective of the coevolution of host galaxies and SMBHs. For this purpose, we compare our results with the theoretical predictions of a coevolution scenario of the galactic bulges and SMBHs. 

\subsection{A Coevolution Scheme for SMBHs and Galactic Bulges}
Recently, KUM03 have constructed a physical model for a coevolution of a QSO BH and an early-type host galaxy. This model is based on the radiation drag incorporating the realistic chemical evolution that reproduces the color-magnitude relation of present-day bulge. Here, we briefly review the essence of their model as a pre-arrangement of the following discussions. In their model, they used an evolutionary spectral synthesis code `PEGASE' (Fioc \& Rocca-Volmerange 1997), in order to treat the realistic chemical evolution of the galactic bulges. 
According to results of KUM03, after a galactic wind epoch $t_{\rm w}$, the bolometric luminosity is shifted from the host-dominant phase to the AGN-dominant phase (the QSO phase) at the transition time $t_{\rm crit}$. The former phase ($t_{\rm w} < t < t_{\rm crit}$) corresponds to the early stage of a growing BH, because the mass accretion rate during this phase is so high that the mass growth of BH is significant. They defined this phase as a ``proto-QSO". The proto-QSO phase is proceeded by an optically thick phase before the galactic wind, which would correspond to a classical ULIRG. In this phase, they predicted that the BH is much smaller than the QSO phase. After the AGN luminosity exhibits a peak at $t_{\rm cross}$, it fades out abruptly because almost all of the matter around BH has fallen onto the central BH. The fading nucleus could be a low luminosity AGN (LLAGN).

\subsection{Are Type I ULIRGs Missing Link Between ULIRGs and QSOs ?}
By using KUM03 model, we predict the evolution of $M_{\rm R}({\rm bulge})$ and $M_{\rm BH}$ for the different masses of bulges in Figure 4. 
Here, we calculate the evolutional tracks in the $M_{\rm R}$-$M_{\rm BH}$ diagram for 4 different masses of bulges ($M_{\rm g0}=10^{10},\, 10^{11},\, 10^{12},\, {\rm and}\, 2\times 10^{12}M_{\odot}$), where $M_{\rm g0}$ is the initial gas mass in galactic bulges. In this figure, we assume a Salpeter-type initial mass function (IMF) to be $\phi = dn/d\log m_{*}=A(m_{*}/M_{\odot})^{-1.35}$ for a mass range of [$0.1M_{\odot}$, $60M_{\odot}$]. 
The star formation rate (SFR) per unit mass at time $t$, $C(t)$, is assumed to be proportional to the gas mass fraction $f_{\rm g}$, $C(t)=kf_{\rm g}$ at $t < t_{\rm w}$ and at $t \geq t_{\rm w}$ $C(t)=0$, where $k$ is the constant rate coefficient. In Table 4, we summarize the model parameters ($M_{\rm g0}$, $k$, $t_{\rm w}$, $t_{\rm crit}$, and $t_{\rm cross}$). \footnote{As for the effect of star formation history, if the mass range and slope of IMF and the SFR are changed to satisfy the spectrophotometric properties of galactic bulges, the final BH mass is altered by a factor of $\pm 50\%$ (Kawakatu \& Umemura 2004).} The evolutions of different stellar masses of bulges proceed from left (smaller BH) to right (larger BH) in Figure 4. The black horizontal arrow denotes the optical extinction effect for the BLR of type I ULIRGs.

\vspace{5mm}
\epsfxsize=8cm 
\epsfbox{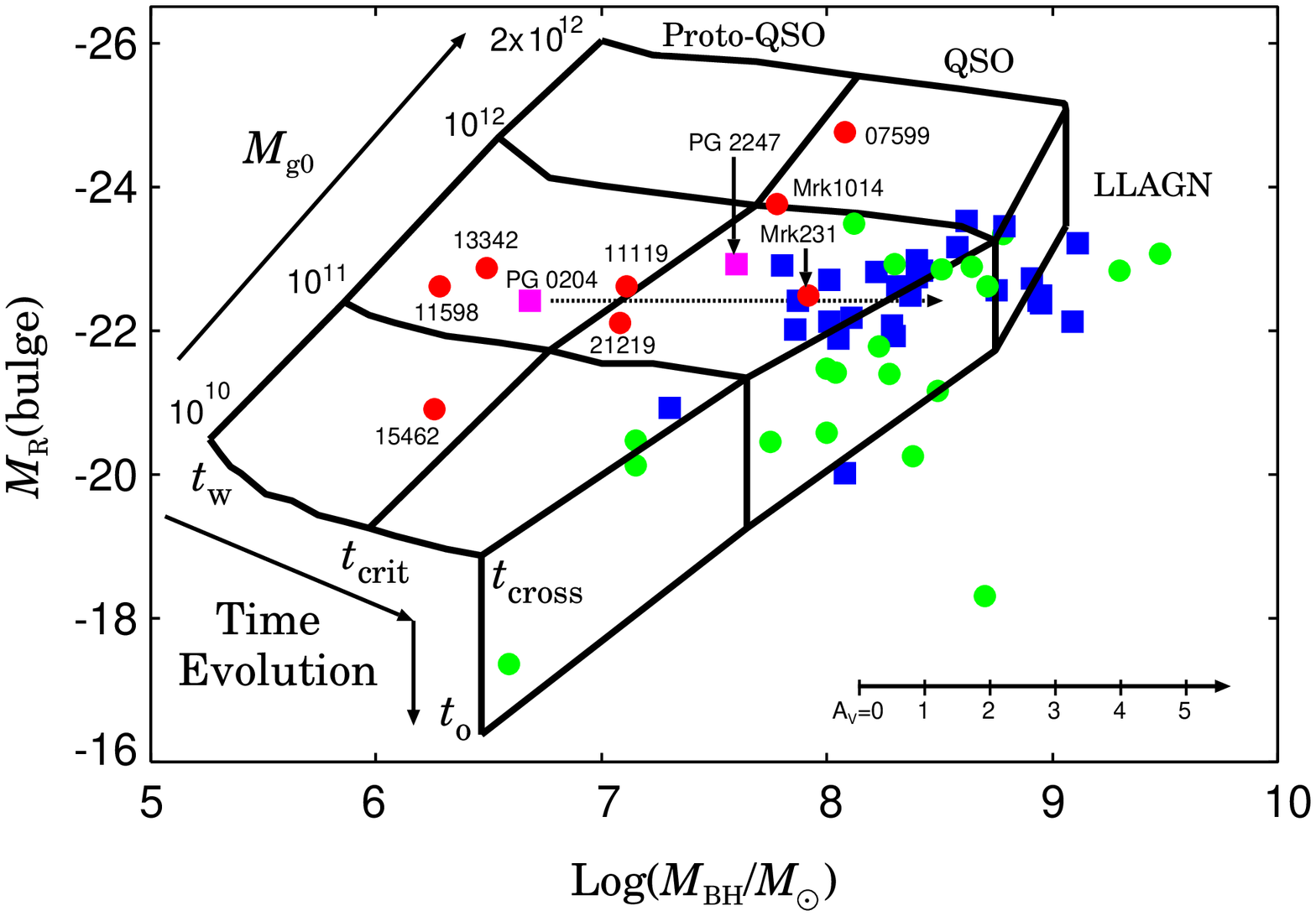}
\figcaption
{
Absolute $R$-band bulge magnitude versus the black hole mass for 8 type I ULIRGs (red circles), 27 QSOs (blue squares), 2 ``unusual" QSOs (magenta squares) and 20 elliptical galaxies (green circles). The dashed arrow denotes the BH mass if we use the broad line width with 8000 km/s. 
The black horizontal arrow shows the optical extinction for the BLR of type I ULIRGs. The grids represent the prediction based on the radiation drag model for different mass of bulges ($M_{\rm g0}=10^{10},\, 10^{11},\, 10^{12},\, {\rm and}\, 2\times 10^{12}M_{\odot}$), where $M_{\rm g0}$ is the initial gas mass in galactic bulges. Each evolutions shift from left (smaller BH) to right (larger BH) in this figure. $t_{\rm w}$ is the galactic wind time scale. $t_{\rm crit}$ is the time when the luminosity of bulges is equal to that of AGNs. $t_{\rm cross}$ is defined as the time when almost all of the matter around BH has fallen onto the central BH. $t_{0}$ denotes the final stage of the galaxy evolution.
The host luminosity dominant phase (proto-QSOs) correspond to the region ($t_{\rm w} < t < t_{\rm crit}$). The AGN luminosity dominant phase (QSOs) correspond to the area ($t_{\rm crit} < t < t_{\rm cross}$). We call the phase ($t_{\rm cross} < t < t_{0}$) a low luminosity AGN (LLAGN).
}
\vspace{5mm}

As seen in Figure 4, it is found that most type I ULIRGs are located near the proto-QSO phase. It would indicate that type I ULIRGs are the early phase of BH growth within younger bulge and then they evolve into QSOs. Namely, it would suggest that type I ULIRGs are the missing link between ULIRGs and QSOs. In addition, we also find that two ``unusual" QSOs (PG 0204+292 and PG 2247+140) are located near the region of the proto-QSOs. In fact, these ``unusual" QSOs have a narrower ${\rm H}\beta$ line width ($< 2500{\rm km/s}$) among the samples of PG QSOs (McLure \& Dunlop 2001). Thus, these may correspond to the objects on the way of evolving into QSOs. However, as for the PG 0204+140, we should comment that it has not only a strong narrow component, but also a very broad line width (about 8,000 km/s: see Figure 6 in McLure \& Dunlop 2001). If we estimate its BH mass by using larger FWHM, PG 0204+140 has the BH mass with $3\times 10^{8}M_{\odot}$ (the dashed arrow). If this is case, it is located near the region of normal QSOs in Figure 4. Additionally, our model can explain the difference from QSOs and elliptical galaxies as the evolution of host galaxies with nearly same BH mass. Moreover, according to KUM03, it would be expected that parent galaxies of type I ULIRGs are the starburst galaxies, in which there is any AGN activity. This corresponds classical ULIRGs. In this phase, their BH mass would be much smaller than the BH mass of type I ULIRGs. 
To sum up, we found that classical ULIRGs---type I ULIRGs---QSOs---ellipticals would be explained by the evolution sequence of the spheroidal systems. However, these are some discrepancy between the predictions of KUM03 model and our results for the type I ULIRGs. In KUM03, they predicted that a proto-QSO is {\it an optically-thin and the total luminosity is still dominated by the host components}. On the other hands, as for type I ULIRGs we revealed the followings; (1) The AGNs powers the optical emission rather than the stellar components (see $\S 3$). It might imply that AGNs contribute the energy source of FIR luminosity. (2) Larger FIR luminosity than the Eddington luminosity indicates the optical depth of their hosts is still thick ($\S4.2$). It is useful to consider the reasons of two disagreements. Hence, we will discuss in next sections.

\subsubsection{Origin of the FIR luminosity}
In KUM03 model, they assume the AGN luminosity to be the Eddington one. 
But, recent optical and X-ray observations have suggested that almost of all type I ULIRGs would be the high mass accretion phases (Anabuki 2004; Hao et al. 2005). If the accretion rate onto a BH is the super-Edddington ($\dot{M}>\dot{M}_{\rm E}$), then the solution of accretion is optically-thick and called a {\it slim-disk} (e.g., Abramowicz et al. 1998), where the Eddinton accretion rate $\dot{M}_{\rm E}=L_{\rm E}/c^{2}$ (e.g., Kato, Mineshige \& Fukue 1998). In this type of disk, the photon-trapping effect (e.g., Katz 1977; Begelman 1978) in the accretion flow plays an important role. In fact, the BH accretion luminosity can achieve up to $\approx 10L_{\rm E}$ (Ohsuga et al. 2002, 2003, and 2005). Hence, by considering such a super-Eddington accretion in the type I ULIRGs, it is expected that the AGN luminosity become larger than the host luminosity, and then this may be consistent with the observational result. If this is the case, moreover, then it turns out that the FIR luminosity of type I ULIRGs can be explained by only the BH accretion luminosity (see thin solid line $L_{\rm Fir}/L_{\rm Edd}=10$ in Figure 3). On the other hands, the recent high resolution multiwavelength observations have indicated that the vast majority of ULIRGs are strongly interacting or merging galaxies (e.g., Clements et al. 1996; Murphy et al. 1996; Veilleux, Kim, \& Sanders 2002). Thus, such a interaction or merger may cause a starburst (e.g., Larson \& Tinsley 1978; Noguchi 1988; Mihos \& Hernquist 1994; Sanders \& Mirabel 1996), and thus a starburst phenomenon may contribute to the FIR luminosity of the type I ULIRGs. However, there's plenty of scope for discussion whether the interactions or mergers trigger the starbursts as Bergvall et al. (2003) claimed that the interactions and mergers do not trigger the starburst. Nevertheless, it is worth emphasizing again that the AGN luminosity only can produce the FIR luminosity of type I ULIRGs if the accretion rate is $\dot{M}>\dot{M}_{\rm E}$. This would impact on one of the important issues for ULIRGs, that is, whether the dominant energy sources of ULIRGs are the dust-obscured AGNs or starbursts. 

\subsubsection{Obscuring problem}
KUM03 was based on the monolithic model for a coevolution of a SMBH and a galaxy. As we mentioned in $\S 5.2.1.$, however, almost all ULIRGs have undergone the mergers or the interactions, and also a AGN phenomenon appears at the final merging stage (e.g., Clements et al. 1996; Kim, Veilleux, \& Sanders 1998; Wu et al. 1998; Zheng et al. 1999; Canalizo \& Stockton 2001; Cui et al. 2001) Thus, the interactions or mergers would play a significant role in order to elucidate the physical link between ULIRGs and QSOs. If dusty galaxies interact or merger with optically thin proto-QSOs, then the dusty gas may be winded up and then the dust-enshrouded AGNs may form. The objects may observe as type I ULIRGs. After the dusty gas would be swept away by the multiple supernovae and AGN feedbacks (e.g., Granato et al. 2004), type I ULIRGs would evolve into the optically thin QSOs. As the other possibility, we may consider that type I ULIRGs may be the on the way to blowing away the dusty gas even within the monolithic scenario. Indeed, the duration of sweeping away the dusty gas is finite, so that the optical depth of bulges would be mildly optically thick during this phase, though we supposed that the galactic wind sweeps the dusty gas in an instant in Figure 4. Thus, in order to understand the physical properties of type I ULIRGs in details, we will elucidate the wind effects and interactions or mergers with the dusty galaxies by using the sophisticated simulations in the future works.

\section{Conclusions}
We have investigated whether type I ULIRGs are really in the transition stage from ULIRGs to QSOs, using the data of infrared, optical and X-ray observations. To reveal this issue, we compare the BH mass, the bulge luminosity and the FIR luminosity among type I ULIRGs, QSOs and elliptical galaxies. Our main conclusions are:

\begin{enumerate}
\item The type I ULIRGs have systematically smaller BH masses in spite of the comparable bulge luminosity relative to QSOs. 

\item The far-infrared luminosity of most type I ULIRGs is larger than the Eddington luminosity. 

\item We have shown that our results do not change significantly for 3 type I ULIRGs that we can evaluate the effects of the visual extinction. 
In addition, we have examined the effect of uncertainty of BH mass measurements and our sample bias, so that we found that our results do not alter drastically.

\item From the X-ray properties of type I ULIRGs, which are similar to those of the narrow line Seyfert 1 galaxies (Anabuki 2004), these would indicate that AGNs with a high mass accretion rate exist in the type I ULIRGs.

\item Based on all of these findings (1-4), it would be a natural interpretation that the type I ULIRGs are the early phase of BH growth, namely the missing link between ULIRGs and QSOs. Moreover, by comparing our results with a theoretical model of a coevolution scenario of a QSO BH and a galactic bulge, we have found that this explanation would be reasonable. 

\item The AGN luminosity only can produce the FIR luminosity of type I ULIRGs if we consider the super-Eddington accretion. This result would impact on the origin of the infrared luminosity in ULIRGs.

\end{enumerate}

\acknowledgments
The authors thank the anonymous for his/her constructive comments and suggestions. The analysis has been made with computational facilities at Center for Computational Sciences in University of Tsukuba. We thank Dr. X. Z. Zheng for friendly supplying the observed spectra of type I ULIRGs. 
We are grateful to M. Akiyama, M. Imanishi, K. Ohsuga and Y. Terashima for useful comments and discussions. N. K. also acknowledges the Italian Ministero dell'Istruzione dell'Universit\'{a} e della Ricerca (MIUR) and the Istituto Nazionale di Astrofisica (INAF) for financial support. N. A. and T. N. thank financial support from the Japan Society for the Promotion of Science (JSPS) through the JSPS Reserch Fellows. This work is supported in part by a grant-in-aid for scientific research from the Ministry of Education, Science, Culture, Sports and Technology in Japan for 16002003 (M.U.).


\clearpage

\begin{table}[t]
\begin{center}
Table 1. The Variable Physical Parameters \\[3mm]
{\scriptsize
\begin{tabular}{lcccc}
\hline \hline
Name & $\log({M_{\rm BH}/M_{\odot}})$ & $M_{\rm R}(\rm host)$ & $\log({L_{\rm FIR}/L_{\odot}})$ & references \\
(1) & (2) & (3) & (4) & (5) \\
\hline
& &Type I ULIRGs&  \\
\hline
IRAS F01572+0009 (Mrk 1014)& 7.78 & -23.76 & 12.30 & 1,2\\
IRAS F07599+6508 &  8.08 & -24.68 & 12.10 & 1,2\\
IRAS F11119+3257 &  7.11 & -22.61 & 12.29 & 1,2\\
IRAS Z11598-0112 & 6.28 & -22.61 & 11.56  &1,2\\
IRAS F12540+5708 (Mrk 231) & 7.92 & -22.49 & 12.25 & 1,2\\ 
IRAS F13342+3932 & 6.49 & -22.87 & 12.14 & 1,2\\
IRAS F15462-0450 & 6.26 & -20.91 & 12.00 & 1,2\\
IRAS F21219-1757 & 7.08 & -22.11 & 11.67 & 1,2\\
\hline 
& &PG QSOs&  \\
\hline 
PG 0137+012 & 8.58 & -23.16 & --- & 3\\
PG 0736+017 & 8.01 & -22.70 & --- & 3\\
PG 1004+130 & 9.11 & -23.22 & $<$ 11.76 &3,7 \\
PG 1020-103 & 8.37 & -22.48 & ---  & 3\\
PG 1217+023 & 8.42 & -22.83 & --- & 3\\
PG 1226+023 & 8.62 & -23.52 & 12.12 & 4,5\\
PG 1302-102 & 8.31 & -22.62 & 12.09 & 4,6\\
PG 1545+210 & 8.94 & -22.42 & $<$ 11.38 & 4,7\\
PG 2135-147 & 8.95 & -22.62 & --- & 3\\
PG 2141+175 & 8.75 & -22.92 & --- & 3\\
PG 2247+140 & 7.60 & -22.92 & --- & 3\\
PG 2349-014 & 8.79 & -23.52 & $<$ 11.57 & 3,7\\
PG 2355-082 & 8.40 & -22.72 & --- & 3\\
PG 0052+251 & 8.29 & -22.12 & --- & 3\\
PG 0054+144 & 8.91 & -22.72 & --- & 3\\
PG 0204+292 & 6.68 & -22.42 & --- & 3\\
PG 0205+024 & 7.86 & -20.22 & --- & 3\\
PG 0244+194 & 8.05 & -21.89 & --- & 3\\
PG 0923+201 & 8.95 & -22.37 & $<$ 11.91 & 3,7\\
PG 0953+414 & 8.40 & -21.98 & --- & 3\\
PG 1012+008 & 7.80 & -21.90 & $<$ 11.48 & 3,7\\
PG 1029-140 & 9.09 & -22.12 & --- & 4\\
PG 1116+215 & 8.22 & -22.82 & $<$ 11.52 & 4,5\\
PG 1202+281 & 8.30 & -21.92 & $<$ 11.38 & 4\\
PG 1307+085 & 7.86 & -22.02 & 11.32 & 4,6\\
PG 1309+355 & 8.01 & -22.12 & $<$ 11.41 & 4,5\\
PG 1402+261 & 7.30 & -20.92 & 11.39 & 4,5\\
PG 1444+407 & 8.08 & -22.02 & 11.57 & 4,5\\
PG 1635+119 & 8.11 & -22.17 & --- & 3\\
\hline
& & Elliptical Galaxies& \\
\hline
NGC 821 & 7.57 & -21.94 & ---  & 8\\
NGC 2778 & 7.15 & -20.12 & --- & 8\\
NGC 3377 & 8.00 & -20.58 & --- & 8\\
NGC 3608 & 8.28 & -21.39 & --- & 8\\
NGC 4291 & 8.49 & -21.16 & --- & 8\\
NGC 4473 & 8.04 & -21.42 & --- & 8\\
NGC 4564 & 7.75 & -20.45 & --- & 8\\
NGC 4649 & 9.30 & -22.83 & --- & 8\\
NGC 4697 & 8.23 & -21.77 & --- & 8\\
NGC 5845 & 8.38 & -20.25 & --- & 8\\
M 32 & 6.59 & -17.36 & --- & 9\\
NGC 3379 & 8.00 & -21.47 & --- & 9\\
NGC 4486B & 8.70 & -18.30 & --- & 9\\
NGC 4742 & 7.15 & -20.47 & --- & 9\\
NGC 4261 & 8.71 & -22.62 & --- & 9\\
NGC 4374 & 8.64 & -22.89 & --- & 9 \\
M87  & 9.48 & -23.06 & --- & 9 \\
NGC 6251 & 8.78 & -23.34 & --- & 9 \\
NGC 7052 & 8.51 & -22.84 & --- & 9 \\
IC 1459  & 8.30 & -22.92 & --- & 9 \\
\hline
\end{tabular}
}
\noindent
\end{center}
{\scriptsize Note.---Col.(1):source name; col.(2) BH mass; col.(3) $R$-band absolute magnitude of host galaxies (mag); col.(4) far-infrared luminosity References---(1) Zheng et al. 2002; (2) Veilleux et al. 2002; (3) Dunlop et al. 2003; (4) McLure \& Dunlop 2001; (5) Haas et al. 2003; (6) Haas et al. 2000; (7) Sanders et al. 1989; (8) Gebhardt et al. 2003; (9) Kormendy \& Gebhardt 2003
}
\end{table}

\begin{table}[t]
\begin{center}
Table 2. The Basic Physical Parameters of Type I ULIRGs\\[3mm]
{\scriptsize
\begin{tabular}{ccccccc}
\hline \hline
IRAS & z & $F_{\lambda} (5100(1+z){\rm \AA})_{\rm obs}$ & FWHM $({\rm H}{\beta})$ & $\lambda L_{\lambda}(5100{\rm \AA})_{\rm rest}$ & $F_{2-10(1+z){\rm kev}}/{F_{5100(1+z){\rm \AA}}}$ \\
(1) & (2) & (3) & (4) & (5) & (6) \\
\hline
IRAS F01572+0009 (Mrk 1014)&  0.163 & $1.41\times 10^{-15}$ & 2100 &$4.6\times 10^{44}$ & $5.3\times 10^{3}$\\
IRAS F07599+6508 &  0.148  &  $1.50\times 10^{-15}$ & 3150 & $3.9\times 10^{44}$ & $9.3\times 10^{1}$ \\
IRAS F11119+3257 &  0.189  & $1.45\times 10^{-16}$ & 1980: & $6.7\times 10^{43}$ & $9.0\times 10^{4}$\\
IRAS Z11598-0112 & 0.151  & $1.72\times 10^{-16}$ & 820 & $4.7\times 10^{43}$ & $1.9\times 10^{4}$\\
IRAS F12540+5708 (Mrk 231) & 0.042  & $1.25\times 10^{-14}$ & 3130 & $2.3\times 10^{44}$ & $3.4\times 10^{2}$\\ 
IRAS F13342+3932 & 0.179  & $1.44\times 10^{-16}$ & 980 & $5.6\times 10^{43}$ & ---  \\
IRAS F15462-0450 & 0.101  & $7.40\times 10^{-17}$ & 1460 & $8.6\times 10^{42}$ & --- \\
IRAS F21219-1757 & 0.113  & $3.24\times 10^{-16}$ & 2080 & $8.0\times 10^{43}$ & --- \\
\hline
\end{tabular}
}
\noindent
\end{center}
{\scriptsize Note.--- The prefix of the object name indicates the origin of IRAS fluxes. ``F'' refers to the IRAS Faint Source Catalogue, and ``Z'' means the Faint Source Reject File. The uncertainty on the FWHM is typically of order $10\%$; colons indicates values with a relative uncertainty of $20\%$ (Zheng et al. 2002). 
Col.(1):source name; col.(2): redshift; col.(3):observed flux at $5100(1+z){\rm \AA}$ [erg s$^{-1}$ cm$^{-2}$ ${\rm \AA}^{-1}$]; col.(4) FWHM of ${\rm H}{\beta}$ emission profile [km s$^{-1}$]; col.(5): continume luminosity [erg s$^{-1}$] at rest frame; col.(6): Flux ratio $F_{2-10(1+z){\rm keV}}/{F_{5100(1+z){\rm \AA}}}$, $F_{2-10{\rm keV}(1+z)}$ is unit of [erg s$^{-1}$ cm$^{-2}$ ${\rm \AA}^{-1}$] \\
References.--- (1)-(5): Zheng et al. (2002) and their original spectra. (6) :Grorge et al. (2000)
}
\end{table}

\begin{table}[t]
\begin{center}
Table 3. The Parameters for the Extinction of Type I ULIRGs\\[3mm]
{\scriptsize
\begin{tabular}{cccccccc}
\hline \hline
IRAS & $N_{\rm H}[{\rm cm}^{-2}]$ & $A_{\rm V} [{\rm mag}]$ & $A_{\rm V}^{\rm galactic} [{\rm mag}]$ & references \\
(1) & (2) & (3) & (4) & (5) \\
\hline
IRAS F01572+0009 (Mrk 1014)& $< 2.6\times 10^{20}$ & $<$ 0.012 & 0.087 & 1 \\
IRAS F07599+6508 & $> 10^{24}$ &  --- & 0.16 & 2\\
IRAS F11119+3257 & $ 1.2\times 10^{22}$ & 0.57 & 0.159 & 1 \\
IRAS Z11598-0112 & $ 6\times 10^{20}$ & 0.029 & 0.066 & 1\\
IRAS F12540+5708 (Mrk 231) & $ > 10^{24}$ & --- & 0.078 & 3\\ 
IRAS F13342+3932 & --- &  --- & 0.012 & --- \\
IRAS F15462-0450 & --- &  --- & 0.65 & --- \\
IRAS F21219-1757 & --- &  --- & 0.21 & --- \\
\hline
\end{tabular}
}
\noindent
\end{center}
{\scriptsize Note.--- 
Col.(1):source name; col.(2): column density; col.(3): internal optical extinctions derived from column demsity. The value is evaluated by $A_{\rm V}/N_{\rm H}=4.8^{+14.1}_{-3.6}\times 10^{-23}{\rm mag\, cm^{-2}}$ ($1\sigma$ dispersion) derived from AGN observation (Maiolino et al. 2001b); col.(4) Galactic extinction at visual band \\
References.--- (1) Anabuki 2004; (2) Braito et al. 2004; (3) Imanishi \& Terashima 2004
}
\end{table}

\begin{table}[t]
\begin{center}
Table 4. The Basic Model Parameters (KUM03 Model)\\[3mm]
{\scriptsize
\begin{tabular}{ccccc}
\hline \hline
$M_{\rm g0}[M_{\odot}]$ & $k[{\rm Gyr}^{-1}]$ & $t_{\rm w}[{\rm yr}]$ & 
$t_{\rm crit}[{\rm yr}]$ & $t_{\rm cross}[{\rm yr}]$ \\
(1) & (2) & (3) & (4) & (5) \\
\hline
$10^{10}$ & 14.7 & $1.3\times 10^{8}$ & $4.5\times 10^{8}$ & $7\times 10^{8}$ \\ 
$10^{11}$ & 11.3 & $3.5\times 10^{8}$ & $8\times 10^{8}$ & $1.2\times 10^{9}$ \\ 
$10^{12}$ & 8.6 & $7\times 10^{8}$ & $1.2\times 10^{9}$ & $1.8\times 10^{9}$ \\ 
$2\times 10^{12}$ & 8.1 & $8\times 10^{8}$ & $1.4\times 10^{9}$ & $2\times 10^{9}$ \\ 

\hline
\end{tabular}
}
\noindent
\end{center}
{\scriptsize Note.--- (1) $M_{\rm g0}$ is the initial gas mass in galactic bulges. (2) $k$ is the constant rate coefficient for SFR. (3) $t_{\rm w}$ is the galactic wind time scale. (4) $t_{\rm crit}$ is the transition time from the host luminosity-dominat phase to the AGN luminosity dominant phase. (5) $t_{\rm cross}$ is the time when the BH growth stops. As for these parameters, see also text and the original paper (KUM03).
}
\end{table}

\end{document}